\documentclass[twocolumn,showpacs,amsmath,amssymb,longbibliography]{revtex4-1}

\usepackage{graphicx}% Include figure files
\usepackage{dcolumn}% Align table columns on decimal point
\usepackage{bm}% bold math
\usepackage{caption}
\usepackage{subcaption}

\begin{document}

\title{Kuramoto Osciilators With Asymmetric Coupling}% Force line breaks with \\

\author{Leonard M. Sander}

 \email{lsander@umich.edu}
\affiliation{%
Department of Physics, 
University of Michigan, Ann Arbor, MI, 48109-1040
}%

\affiliation{
Center for the Study of Complex Systems, 
University of Michigan, Ann Arbor, MI, 48109-1042% with \\
}%

\date{\today}% It is always \today, today,

\begin{abstract}
We study a system of  coupled oscillators of the Sakaguchi-Kuramoto type with interactions including a phase delay. We consider the case of 
a coupling matrix such that oscillators with large natural frequencies drive all slower ones but not the reverse. This scheme is inspired by Hebbian learning in neuroscience. We propose a simple model which is partly solvable analytically and  shows many unusual features. For example, for the case of phase delay $\pi/2$ (a symmetric phase response curve) with symmetric coupling  there is no synchronous behavior. But in the asymmetric coupling case, there are phase-locked states, i.e., states where many oscillators have the same frequency, but with substantial phase differences.
\end{abstract}

%\pacs{XXX}% PACS, the Physics and Astronomy
\maketitle

\section{\label{sec:intro}Introduction}
In this paper, we look at a simplified model of coupled oscillators motivated by treatments of   neuron interaction with learning. A numerical approach to some of the results has recently appeared \cite{Aktay2024} (ASZ). Here, we explore in depth some of the unexpected results that  arose in ASZ  because  they took  the coupling between oscillators to be asymmetric: i.e. oscillator $j$ drove oscillator $i$ with a different strength than $i$ drove $j$. Such an asymmetry is generic in systems of neurons coupled by synapses. In  ASZ it was the result of using a model for learning.

Our work is based on the Sakaguchi-Kuramoto (SK) \cite{Sakaguchi1988}  scheme of $N$ coupled oscillators with a phase delay, $\alpha$:
\begin{equation}
\label{SKeqn}
\dot \theta_i = \omega_i - \sum_j K_{ij}\sin(\theta_i-\theta_j + \alpha)
\end{equation}
Here, $\theta_i(t)$ is the phase of the $i^{th}$ oscillator at time $t$,   $\omega_i$ is its natural (uncoupled) frequency, and $K_{ij}$ is the coupling matrix. SK took the  $K_{ij}$   to be a multiple, $K/N$, of the unit matrix, i.e., all-to-all coupling.

The phase delay, 
$\alpha$, with $ 0 \le \alpha \le \pi/2$, tunes between antisymmetric and symmetric drive of the $i^{th}$ oscillator by the $j^{th}$. For $\alpha=0$ (the original Kuramoto model \cite{Kuramoto1984,Acebron2005}) complete synchrony  of some or all of the oscillators occurs if  $K$ is large enough. For $\alpha = \pi/2$, synchrony is absent for all $K$.  

The change in the nature of the driving function is reminiscent of the change in behavior of neurons according to their phase response curves (PRC) \cite{Aktay2024} . Neurons with Type 1 PRC's act like the $\alpha=\pi/2$ case, and Type 2 are like the $\alpha=0$ case. 
Type 1 neurons cannot synchronize, but Type 2 neurons can.
They are sometimes called integrators and resonators in the neuroscience literature.

In most treatments of the SK model (and its forebearer, the Kuramoto model \cite{Kuramoto1984,Acebron2005}) the coupling constants, $K_{ij}$, are a symmetric matrix. However, \textit{asymmetric} couplings are common in actual neuron systems. In particular, a well-known form of learning, spike-time dependent plasticity \cite{Bi2001,Maistrenko2007}, tends to result in strong synapses when  fast neurons (large $\omega_i$) drive slower ones, but weak synapses in the reverse direction.  

In ASZ it was shown that this case  the general behavior described in the previous paragraph changes completely. In both Type 2 and Type 1 cases, we found frequency locking, namely entrainment of the frequencies of some of the oscillators to the fastest one. The relative phases of the frequency-locked oscillators differ substantially. 

There is a large literature about a form of frequency locking that occurs for \textit{identical} oscillators, i.e. $\omega_i \equiv \omega$ in Eq.~(\ref{SKeqn}); see, for example, \cite{Berner2021} and references therein. In many situations this gives rise to \textit{splay states}, frequency-locked states with equally spaced phases. In our case we take the frequency spacing to be non-zero. In fact, we take it to be a fraction of  $K$ in Eq.~(\ref{SKeqn}) (they have the same units). The resulting relative phases are not equally spaced, but, in the case of Type 1, the spacings do not vanish as the coupling increases, as they would for true synchrony. 

In this paper we explore, both by numerics and analysis, a very simplified version of Eq.~(\ref{SKeqn}), far simpler than the ASZ treatment.  In our toy model, we assume \textit{fully} asymmetric coupling such that fast -to-slow drive is present, but the reverse is completely absent. The toy model has remarkable (and remarkably simple) features that we have been able to explain, in part, analytically using elementary techniques.

\section{Model Definition}
In our toy model we use Eq.~(\ref{SKeqn}) with:
\begin{eqnarray}
\label{couplingmatrix}
K_{ij} & = & K \quad \omega_i <  \omega_j\\  \nonumber
  & = & 0  \quad  \omega_i > \omega_j
\end{eqnarray}
We number the oscillators  $i= 0, ..., N-1$ such that lower indexes correspond to larger $\omega_i$. Further, we suppose that the $\omega$'s are \textit{equally spaced}, and put:

\begin{equation}
 \label{eq:omegas}
\omega_i= \omega_0 - i\Delta, 
\end{equation}
where $\Delta$ is the frequency spacing. Of course, the numerical value of $\omega_0$ plays no real role in the results, since we can shift to a rotating frame. So, without loss of generality, we can take $\omega_0 = N\Delta$.

Now we can divide the equation by $\Delta$ and redefine time as $\tau = \Delta t$. We redefine the 'dot' notation to mean $d/d\tau$. Then Eq.(\ref{SKeqn}) becomes:

\begin{eqnarray}
\label{modeleq}
\dot \theta_i &=&  \Omega_i - k \sum_{j=0}^{i-1} \sin(\theta_i-\theta_j + \alpha) \\  \nonumber
\Omega_i  &=& N, N-1, N-2, ... , 1; \quad i=0,.. ,N-1  \\  \nonumber
k &=& K/\Delta
\end{eqnarray}
Note that these equations are in the form of a recursion relation. For $i=0$, we simply have $\theta_0 = \Omega_0 t + \phi_0$. Next, we use this result to find $\theta_1(t)$. Then $\theta_0(t)$ and $\theta_1(t)$ determine $\theta_2(t)$, and so on. Put another way, the faster oscillators act as an external drive to the slower ones.

In what follows, we will focus on two cases, $\alpha= \pi/2$ and $\alpha= 0$, that is, Type 1 and Type 2 oscillators. In these cases the term inside the sum in Eq.(\ref{modeleq}) becomes $$ \cos(\theta_i-\theta_j ),  \sin(\theta_i-\theta_j ),$$ respectively.

In the original Kuramoto \cite{Kuramoto1984} and SK \cite{Sakaguchi1988} cases, the approach was to let $N \to \infty$ with fixed $K/N$. We are concerned with a different limit here. We suppose $N$ to be finite and vary $k$.

\section {Numerical Results}
In this section, we explore the model numerically for Type 1 and Type 2 oscillators.

\subsection{Type 1:  $\alpha=\pi/2$}
For Type 1 coupling, the equations are:

\begin{eqnarray}
\label{type1eq}
\dot \theta_0 &=& \Omega_0 \\  \nonumber
\dot \theta_1 &=& \Omega_1 - k\cos(\theta_1 -\theta_0)  \\  \nonumber
\dot \theta_2 &=& \Omega_2 - k(\cos(\theta_2 -\theta_0) + \cos(\theta_2-\theta_1)) \\  \nonumber
&.....& \\  \nonumber
\dot \theta_j &=& \Omega_j -k \sum_{j=0}^{i-1} \cos(\theta_i-\theta_j ).
\end{eqnarray} 
We will be interested in the long-time behavior of the phases. We integrated the set of equations in Eq. (\ref{type1eq}), using standard ode integrators in Python.

\begin{figure}[htbp]
\centering

\includegraphics[width=0.22\textwidth]{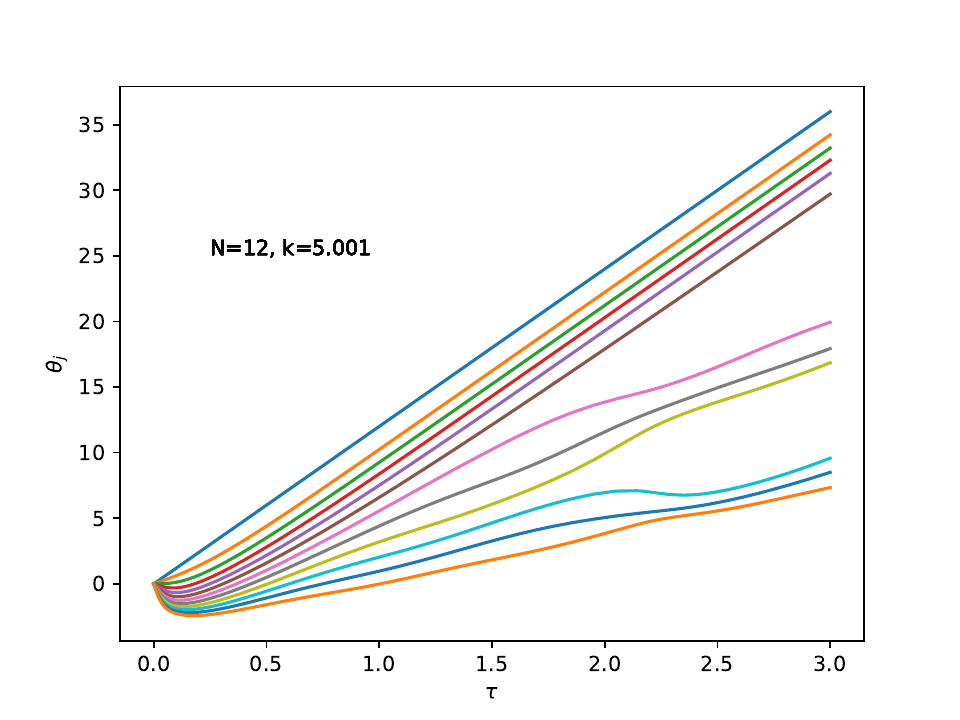}
\includegraphics[width=0.22\textwidth]{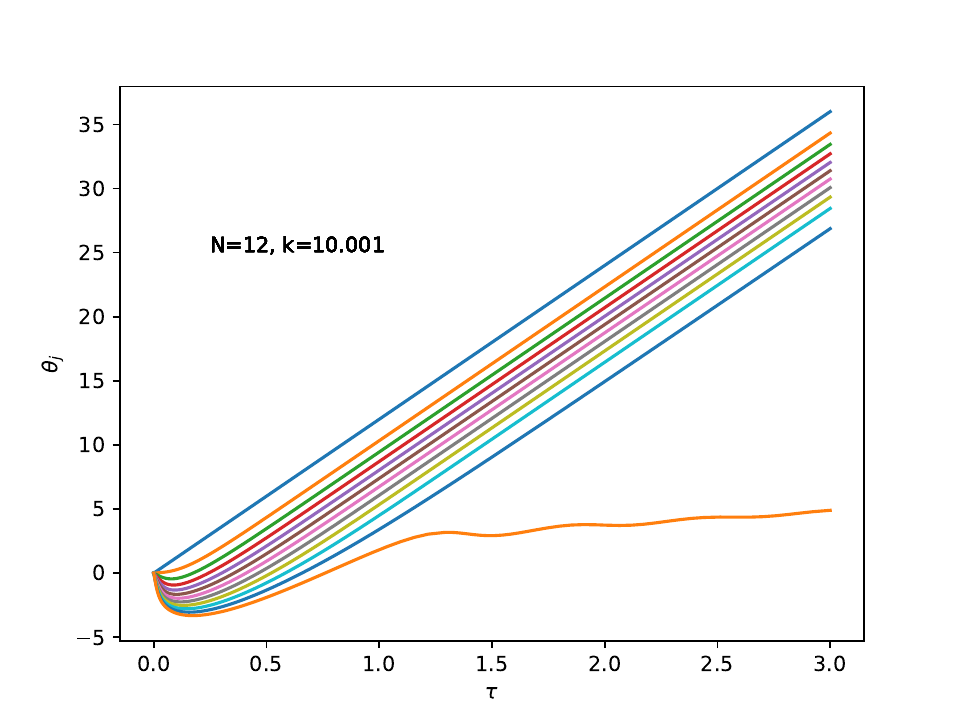}
 \caption{Phase as a function of time for $N=12$ and $k=5^+$ (left) and $10^+$ (right).  The initial phases are all zero. Five oscillators are entrained by the fastest one in the first case, and ten in the second. The lines are the oscillator phases in order from top to bottom.}  
 \label{fig:phases} 
\end{figure}

In Figure (\ref{fig:phases}), we show the results of solving Eq.~ (\ref{type1eq}) for $N=12$ for two values of $k$. The initial phases are all taken to be zero. As $k$ increases we find that a number of oscillators, $N_e$, are entrained to rotate at the fastest frequency. It is plausible that for very small $k$,  $N_e=0$, and that $N_e$ should increase with $k$. Thus  there should be  a series of thresholds where $N_e$ increases by one. So, for very small $k$ there is no entrainment. As $k$ increases (in fact, when $k\ge1$, as we will see), one oscillator follows the leader, $N_e=1$. For larger $k$, there is a point when $N_e=2$, and so on.

What is not evident but seems to be true numerically, is that \textit{the thresholds are equally spaced and are integers}. In fact, our numerical results are that: 
\begin{equation}
\label{Neeq}
\rm{if} \quad  j \le k < j+1, {then} \quad N_e = j,
\end{equation}
or, equivalently, $N_e= \rm{int}(k)$, where int denotes the integer part.
 The data in Figure~(\ref{fig:phases}) is consistent with this proposition.  We will show how this follows from Eq.~(\ref{type1eq}) in a following section.

\begin{figure}[htbp]
\begin{center}
\includegraphics[width=0.5\textwidth]{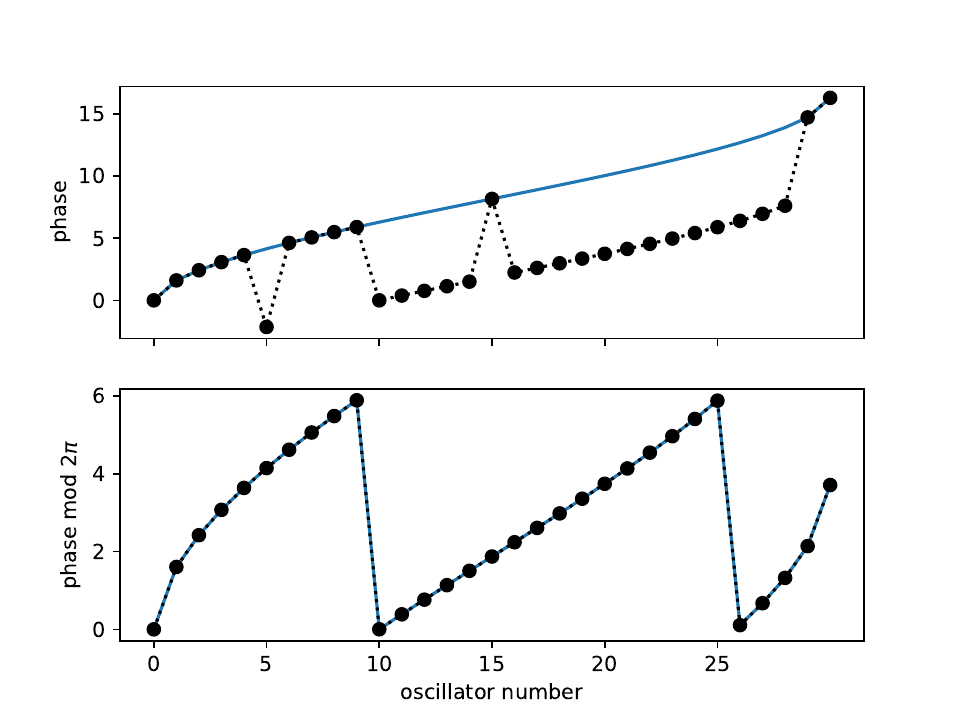}
\caption{\label{fig:phasesmod} 
Plot of $\theta_0(t) - \theta_j(t)$  for $N=32, k=31.001, t=5$ as a function of oscillator number for $j=1,..30$.  Top: blue line, $\theta_i(0) \equiv 0$. Dots, $\theta_i(0)$ chosen at random in $[0,2\pi]$. Bottom panel, same data, modulo $2\pi$.}
\end{center}
\end{figure}

\begin{figure}[htbp]
\begin{center}
\includegraphics[width=0.5\textwidth]{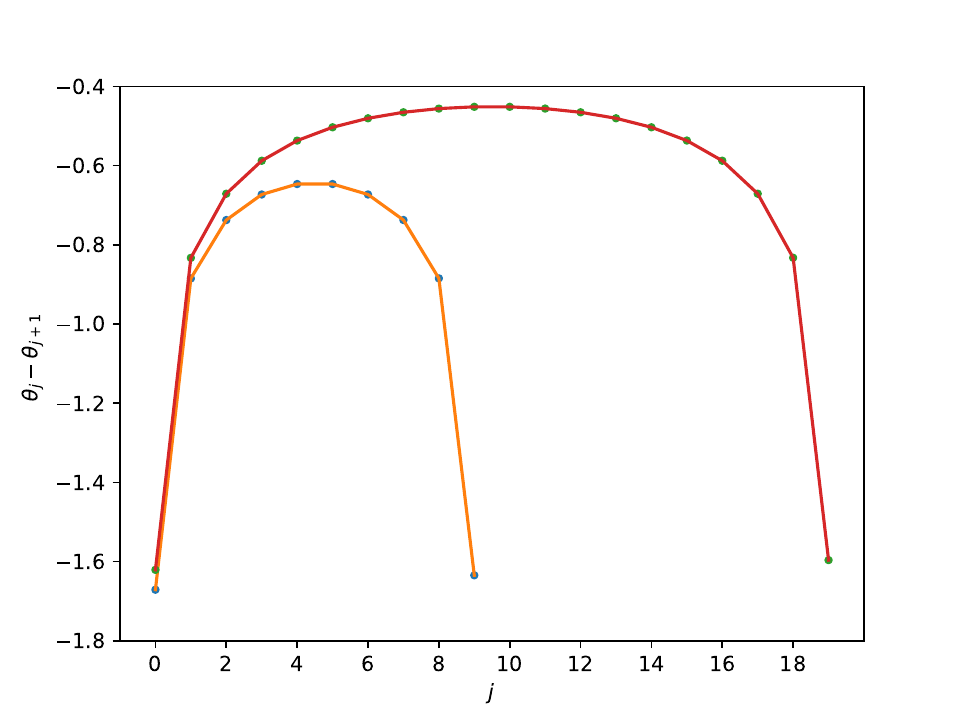}
\caption{\label{fig:phasediffs} 
Plot of $\delta_j = \theta_j - \theta_{j+1}$ at  $t=10$ for $N_e = 10$, lower curve, and $N_e=20$, upper curve.}
\end{center}
\end{figure}

The pattern of phase differences for the entrained oscillators is quite remarkable as well (see Figures~(\ref{fig:phasesmod}) and (\ref{fig:phasediffs})). The first thing to note is that initial phases, $\theta_i(t=0)$, play no real role in the final outcome. If the initial phases are not all zero, the phases in the long-time limit are the exactly the same, modulo $2\pi$. The only effect of the initial conditions is to make some of the oscillators turn a few extra full rotations. That is, the numerical evidence seems to point to the final state being a stable attracting state for the dynamics.

A close examination of Figure~(\ref{fig:phases}) shows that the long-time phases seem to  have a peculiar symmetry property, namely that 
$\theta_0-\theta_1 \approx \theta_{N-2} - \theta_{N-1}$, and $\theta_1-\theta_2 \approx \theta_{N-2}- \theta_{N-3}$, etc. This is reflected in the $s$-shape of the blue line in Figure~(\ref{fig:phasesmod}). We can see this most clearly by plotting:
$$ \delta_j = \theta_j - \theta_{j+1}  \quad j=0,  1,..N_e-1,$$
as in Figure~(\ref{fig:phasediffs}). Note the remarkable symmetric shape of the curve and the apparent similarity of shape for the two different values of $N_e = \rm{int}(k)$.

\begin{figure}[htbp]
\begin{center}
\includegraphics[width=0.5\textwidth]{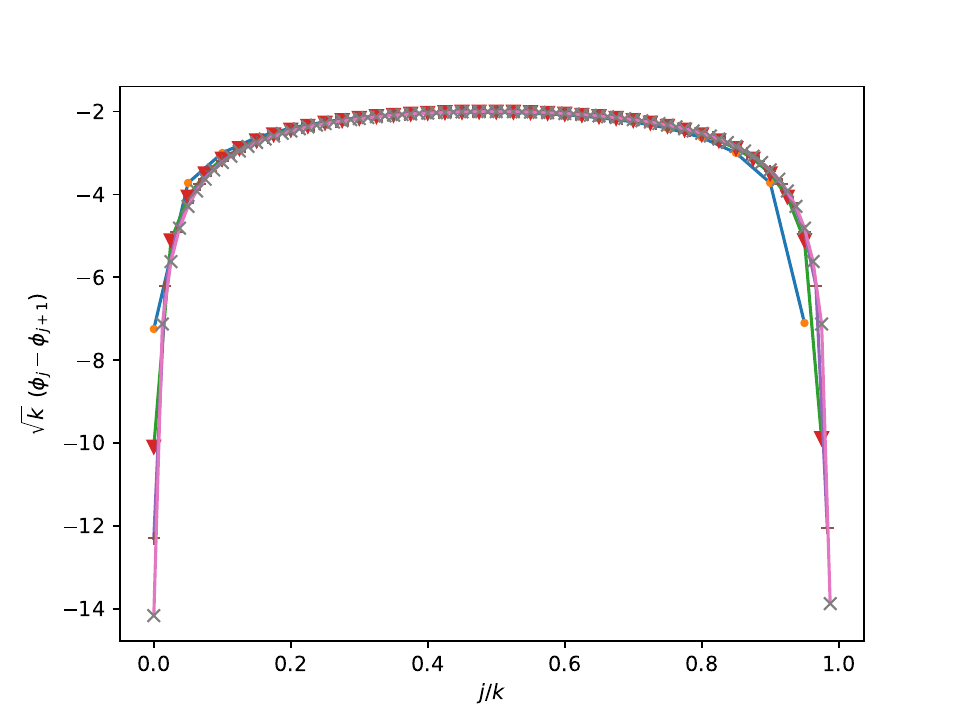}
\caption{\label{fig:diffclps} 
Plot of $\sqrt{k} \delta_j = \sqrt{k}(\theta_j - \theta_{j+1})$ versus $j/k$ at  $t=10$ for $N_e = \rm{int}(k)= 20 (.), 40 (\nabla), 60 (+), 80 (\rm{x}) $.}
\end{center}
\end{figure}

In fact, it is possible to collapse plots of $\delta_j$ into a scaling relation for large  $k$:
\begin{equation}
\label{datacollapseeq}
\delta_j = \frac{1}{\sqrt{k}} F(j/k), \quad j=0, 1, .., N_e.
\end{equation}
The scaling function, $F$, is shown in Figure(\ref{fig:diffclps}) for four values of $k$ at fixed time. This function appears to be symmetric around the value 1/2. The scaling is best in the middle of the figure.

\subsection{Type 2, $\alpha=0$}
Type 2 oscillators are qualitatively similar to Type 1 in the sense that there are thresholds for entrainment, and the slower oscillators are captured one at a time. However, this case is very different quantitatively: the thresholds are much smaller and closer together, and the pattern of phases is quite different.

For Type 1 coupling, the equations are:
\begin{eqnarray}
\label{type2eq}
\dot \theta_0 &=& \Omega_0 \\  \nonumber
\dot \theta_1 &=& \Omega_1 - k\sin(\theta_1 -\theta_0)  \\  \nonumber
\dot \theta_2 &=& \Omega_2 - k(\sin(\theta_2 -\theta_0) + \sin(\theta_2-\theta_1)) \\  \nonumber
&.....& \\  \nonumber
\dot \theta_j &=& \Omega_j -k \sum_{j<i} \sin(\theta_i-\theta_j ).
\end{eqnarray}

\begin{figure}[htbp]
\begin{center}
\includegraphics[width=0.5\textwidth]{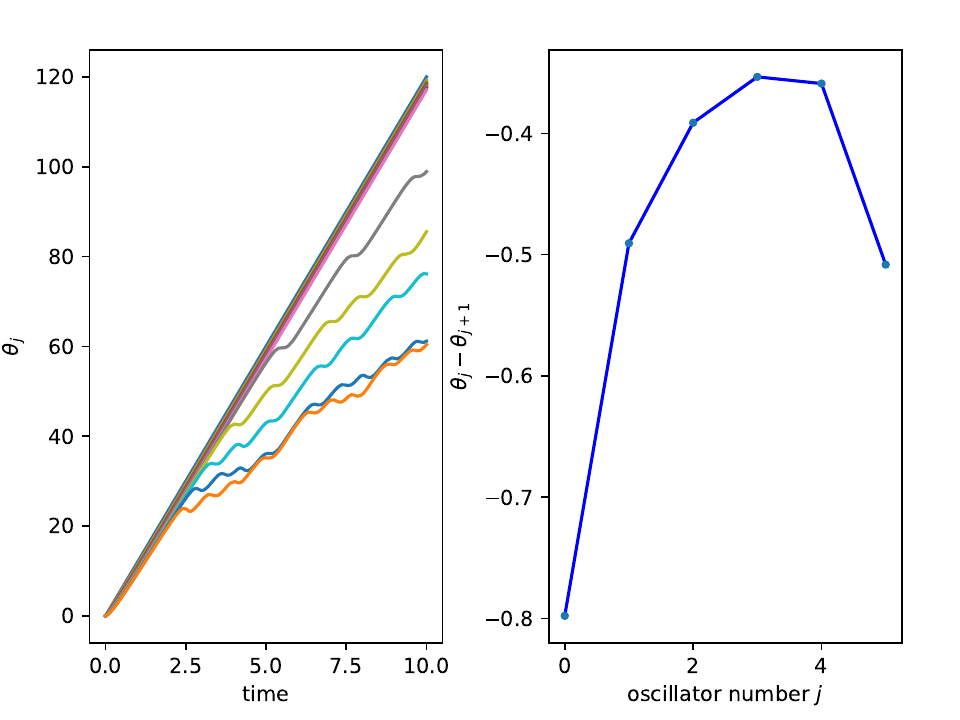}
\includegraphics[width=0.5\textwidth]{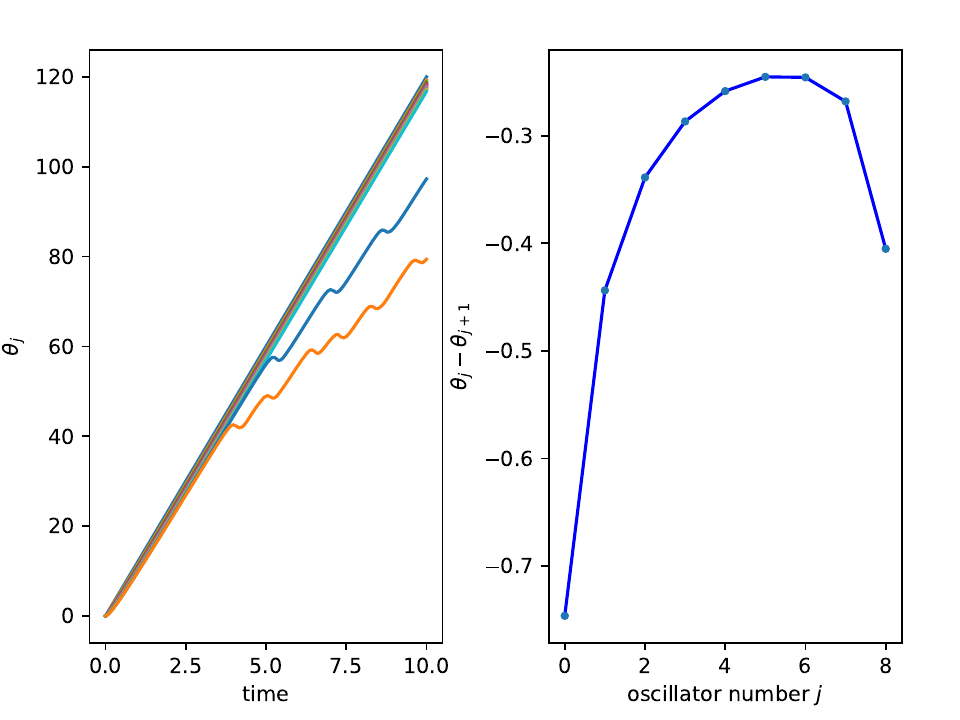}
\caption{\label{fig:phasediffs2} 
Phases versus time, and phase differences for Type 2.  Top: $t=10 $ for $N=12, k=1.397, N_e=6$.
Bottom: $N=12, k=1.473, N_e=9.$}
\end{center}
\end{figure} 

\begin{figure}[htbp]
\begin{center}
\includegraphics[width=0.5\textwidth]{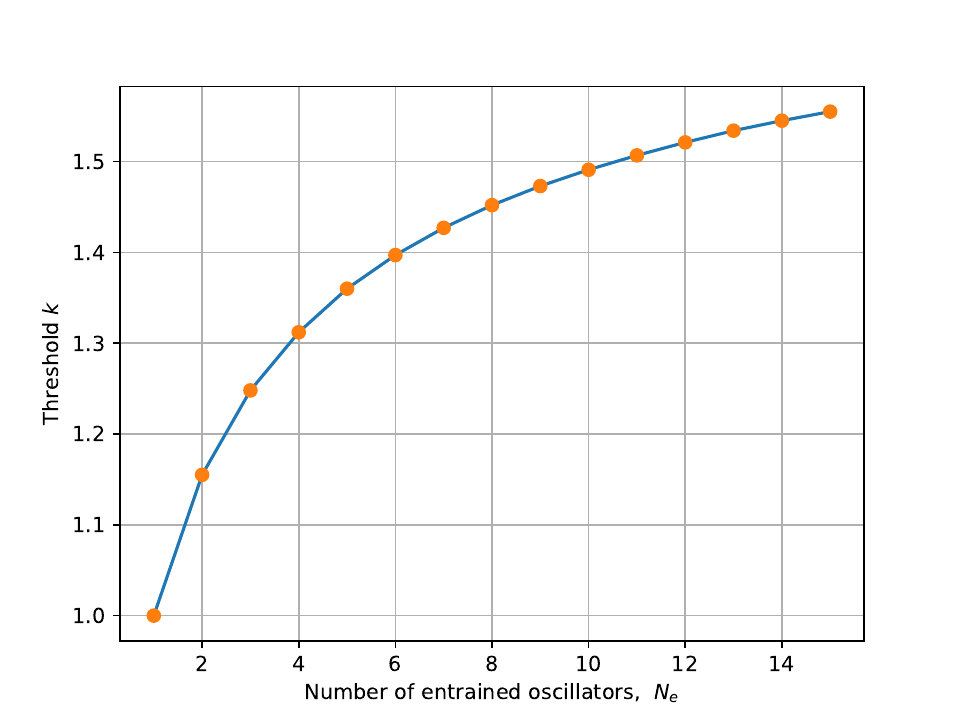}
\caption{\label{fig:thres2} 
Threshold $k$'s for entraining $N_e$ oscillators versus $N_e$ for Type 2 oscillators.}
\end{center}
\end{figure}

In Figure~(\ref{fig:phasediffs2}),  we show examples of the behavior we find by integrating Eqs.~(\ref{type2eq}). The thresholds are $k=1.397, N_e=6$ and $k=1.473, N_e=9$. The lovely symmetry of Figure~(\ref{fig:phasediffs}) is completely absent.
However, in this case, as in the Type 1 case, the phases of the entrained oscillators are independent of initial conditions, modulo $2\pi$.
The thresholds are given in Figure ~(\ref{fig:thres2}). This slowly varying curve is quite different from the simple result of Eq.~(\ref{Neeq}). The curve does not fit very well to any simple power law.

\section{Analytical Treatment}
In this section, we attempt to understand the numerical results  by analyzing Eqs. (\ref{type1eq}) and (\ref{type2eq}). We have a significant result for Type 1, namely a proof of Eq. (\ref{Neeq}). We have made very little progress for Type 2. 

\subsection{Type 1:  $\alpha=\pi/2$}
If there is frequency locking, in Figure~(\ref{fig:phases}) we are led to the following \textit{ansatz} for $i=0,1,...,N_e$:
\begin{equation}
\label{defphi}
\theta_i(t) = \Omega_0 t + \phi_i,
\end{equation}
where $\phi_i$ are the phases of the oscillators relative to the leading one.
The phase of the fastest oscillator, $\phi_0$, is arbitrary.

Inserting this form into Eqs.~(\ref{type1eq}.) we find for $i \le N_e$:
\begin{eqnarray}
\label{type1eqphi}
-\frac{1}{k} &=&\cos(\phi_1 -\phi_0)  \\  \nonumber
-\frac{2}{k} &=& \cos(\phi_2 -\phi_0) + \cos(\phi_2-\phi_1) \\  \nonumber
&.....& \\  \nonumber
-\frac{i}{k} &=&  \sum_{j=0}^{i-1} \cos(\phi_i-\phi_j ).
\end{eqnarray} 
The value of $N_e$ is determined by finding how many of these equations have a solution for a  given $k$.

Clearly, on examining the first line of Eqs.(\ref{type1eqphi}), we see that we need $k \ge 1$ to determine $\phi_1$. Thus, the first threshold is at $k=1$, and, at threshold, $\phi_1=\pi$. (We have put $\phi_0=0$). It is easy to show explicitly that for $k \ge 2$ , we have two equations with solutions, namely: $\phi_1=\pm 2\pi/3, \phi_2= \pm 4\pi/3$, As we will see later, the negative solutions are stable, and the positive ones unstable. For $k$ just above 3, the stable solution is: $\phi_1=\cos^{-1}(-1/3)= -1.911, \phi_2=-\pi$, and $\phi_3= -\pi + \phi_1$. These explicit results, i.e., thresholds at $k=1,2,3$, agree with Eq.~ (\ref{Neeq}). They also have the symmetry property of Figure (\ref{fig:phasediffs}).

\subsubsection{Thresholds}
Let us suppose that the $l^{th}$ equation of the set in Eqs.~(\ref{type1eqphi}) that has a solution. Then for it, and any earlier equation, we can write:

\begin{eqnarray}
\label{threseq}
-l/k &=& \cos(\phi_l-\phi_0) + \cos(\phi_l - \phi_1) + ... \\  \nonumber
   &=& \cos(\phi_l) [ \cos(\phi_0) + \cos(\phi_1) + ..] \\  \nonumber
   & &\quad + \sin(\phi_l) [ \sin(\phi_0) + \sin(\phi_1) + ..] \\  \nonumber
   &=& \cos(\phi_l) \sum_{i=0}^{l-1} \cos(\phi_i) + \sin(\phi_l) \sum_{i=0}^{l-1} \sin(\phi_i) \\  \nonumber
   &\equiv& f_l \cos(\phi_l) + g_l \sin(\phi_l), \nonumber
    \end{eqnarray}
    where:
   \begin{equation}
   \label{deffg}
   f_l = \sum_{i=0}^{l-1} \cos(\phi_i) ; \quad g_l = \sum_{i=0}^{l-1} \sin(\phi_i).
 \end{equation}
 Now introduce new variables, $r_l, \psi_l$ as follows;
 \begin{equation}
 \label{threseq2}
 f_l  =  r_l \cos(\psi_l); \quad g_l = r_l \sin(\psi_l).
 \end{equation}
or, 
 \begin{equation}
 \label{threseq2a} 
 r_l^2 = f_l^2 + g_l^2; \quad \tan(\psi_l) = g_l/f_l. 
 \end{equation}
 The phase angle $\psi_l$ depends on the earlier phases in a complicated way.
 
 Now inserting Eq.~(\ref{threseq2}) into  Eq.~(\ref{threseq}),  we have:
 \begin{equation}
 \label{threseq3}
 -\frac{l}{k r_l} = \cos(\phi_l - \psi_l).
 \end{equation}
 This equation has a solution when $l/kr_l \le 1$.
 
 This is a useful exercise because we can compute $r_l$ from Eqs.~(\ref{type1eqphi}). First we note that:
 \begin{eqnarray}
 \label{rleq}
 r_l^2  &=& f_l^2 + g_l^2 \\ \nonumber
    & = & \left( \sum_{i=0}^{l-1} \cos(\phi_i) \right)^2  + \left( \sum_{i=0}^{l-1} \sin(\phi_i) \right)^2 \\ \nonumber   
    & = & \sum_{i=0}^{l-1}[\cos(\phi_i)^2 + \sin(\phi_i)^2] \\ \nonumber
   & & + 2 \sum_{i=1}^{l-1} \sum_{j=0}^{n-1}[\cos(\phi_i) \cos(\phi_j) + \sin(\phi_i) \sin(\phi_j)]  \\ \nonumber
   &=& l + 2 \sum_{n=1}^{l-1} \sum_{m=0}^{n-1}\cos(\phi_n-\phi_m) 
 \end{eqnarray}
The second term (the cross terms in the squares of the sums in Eq.~(\ref{rleq})) can be evaluated by summing Eqs.~(\ref{type1eqphi}) up to index $l-1$:
\begin{equation}
\label{sumeq}
-\sum_{i=1}^{l-1} \frac{i}{k} =  -\frac{l(l-1)}{2k} = \sum_{i=1}^{l-1} \sum_{j=0}^{i-1} \cos(\phi_i-\phi_j).
\end{equation}

Therefore:
\begin{equation}
\label{rl2}
r_l^2 = l - \frac{l(l-1)}{k} = \frac{l}{k} (k+1-l).
\end{equation}
Now Eq.~(\ref{threseq3}) reads:
\begin{equation}
\label{threseq4}
-\left(\frac{l/k}{k+1-l}\right)^{1/2} = \cos(\phi_l -\psi_l).
\end{equation}

The threshold condition is:
\begin{eqnarray}
\label{threseq5}
\frac{l/k}{k+1-l} &\le& 1 \quad \rm{thus} \\ \nonumber
l &\le& k. \nonumber
\end{eqnarray}
This proves Eq.~(\ref{Neeq}).

\subsubsection{Stability}  
Investigating stability in a coupled system normally requires calculating the eigenvalues of a Jacobian. This case is much simpler because of the recursive nature of Eq.~(\ref{modeleq}). We can use an inductive technique, i.e. assume stability up to $l-1$ and then examine the $l^{th}$ equation. 

To this end we consider the substitution:
$$\theta_l(t) = \Omega_0 t + \phi_l + \delta(t),$$ in Eq.~(\ref{type1eq}). We suppose $\delta << 1$.
We assume the phases are known for $j<l$. Then we have, after a few lines of algebra:
\begin{eqnarray}
\label{stabilityeq1}
\dot \delta &=& k \delta \sum_{j=0}^{l-1}\sin(\phi_l -\phi_j) \\ \nonumber
      &=& k \delta \left( \sin(\phi_l) \sum_{j=0}^{l-1} \cos(\phi_j) - \cos(\phi_l) \sum_{j=0}^{l-1} \sin(\phi_j) \right) \\  \nonumber
      &=& k\delta    \left( f_l \sin(\phi_l) - g_l \cos(\phi_l) \right) \\ \nonumber
      &=& kr_l \delta \sin(\phi_l -\psi_l). \nonumber  
\end{eqnarray}
We have used the definitions of $f_l, g_l$ from Eq.~(\ref{threseq}). The solution is stable if $\sin(\phi_l - \psi_l)$ is negative.

Suppose $k$ is not at any threshold (i.e. is not an integer). Suppose there is stability up to $l-1$. For $l$ there are two solutions to Eq.~(\ref{threseq3}): one is stable, and the other, unstable:
\begin{eqnarray}
\label{stabilityeq2}
\phi_l - \psi_l &=& \pm | \cos^{-1}(-l/kr_l) |. \\ \nonumber
\sin(\phi_l-\psi_l) &=& \pm \sin(| \cos^{-1}(-l/kr_l) |) \nonumber
\end{eqnarray}

\subsection{Type 2, $\alpha=0$}
For Type 2 we have made essentially no progress analytically. We can proceed a bit, and put 
$$\theta_i = \Omega_0 t +\phi_i,$$
as before. We are led to the following:
\begin{equation}
\label{type2eqphi}
-\frac{i}{k} =  \sum_{j=0}^{i-1} \sin(\phi_i-\phi_j ).
\end{equation}

The construction of the analogs to $r_l, f_l, g_l$ in Eq.~(\ref{threseq}) still works, but we are unable to compute $r_l$ in this case. However, we can still  prove that for $l \le N_e$ we have one stable, and one unstable solution for $\phi_l$.

\section{Discussion}

\begin{figure}[htbp]
\begin{center}
\includegraphics[width=0.5\textwidth]{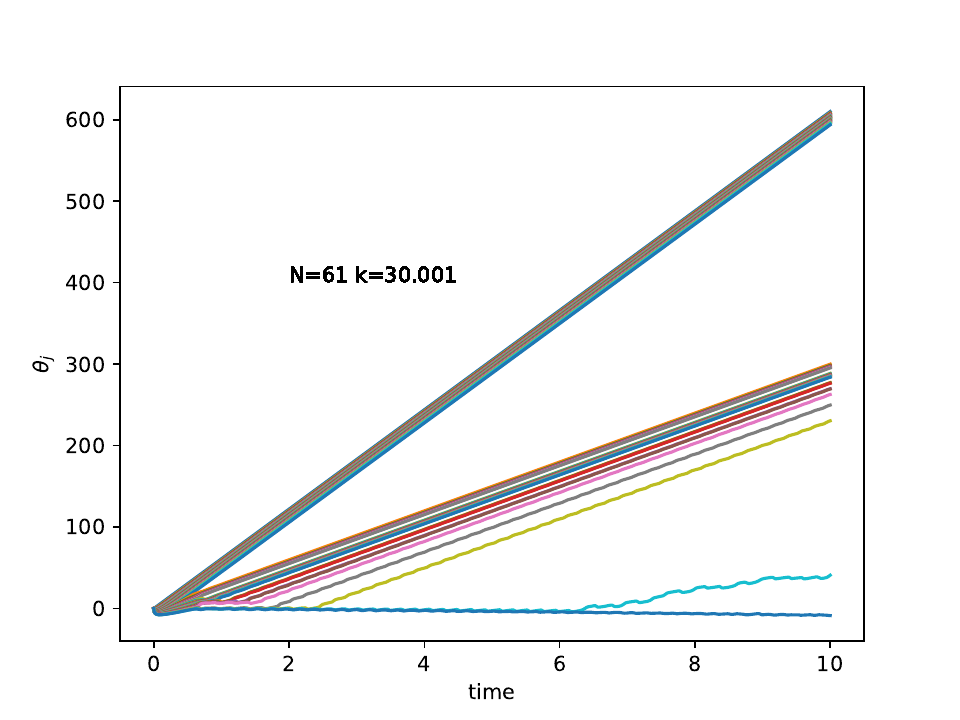}
\caption{\label{fig:phases6130} 
Phases versus time for $N=61, N_e=30$. Note the second phase-locked group.}
\end{center}
\end{figure}

The model we have presented is deceptively simple, and, at least for Type 1, has many unexpected and remarkable features. We have been able to understand the most striking result, $N_e= \rm{int} (k)$, but there is still much to be learned. 

Viewed as a mathematical exercise, the model is quite attractive: it is easy to write down, and  it does not seem to require advanced techniques. The reader is invited to try to derive, for example, the striking symmetry and scaling properties of Figures (\ref{fig:phasediffs}) and (\ref{fig:diffclps}).

Further, there are many extensions that could be made. For example, we have only looked at $\alpha=\pi/2$ and $\alpha =0$. What about the mixed case, e.g. $\alpha = \pi/4$? The numerical results of ASZ seem to indicate that in this case, phase locking persists. Also, suppose $K_{ij}$ is not fully one-directional, but has a symmetric part. Once more, ASZ find echos of the extreme case. 

We have not said anything about the oscillators that are not phase-locked. In Figure (\ref{fig:phases6130}), we show a phenomenon that we often encounter: in addition to the leading group, there can be phase-locking to other frequencies so that there are groups of entrained oscillators.

Whether this model is relevant to actual neuroscience is a much harder question. It is certainly true that in the brain there are many asymmetric connections. After all, axons and dendrites are quite distinct. However, this author is not aware of brain behavior that could be related to this approach, but there certainly could be. In other applications of coupled oscillators there also could be asymmetry so that this kind of model could be relevant.

\begin{acknowledgments}
The author would like to acknowledge useful conversations with Michal Zochowski and Sinan Aktay.
\end{acknowledgments}
\bibliography{asykurabib}% Produces the bibliography via BibTeX.

\begin{thebibliography}{7}%
\makeatletter
\providecommand \@ifxundefined [1]{%
 \@ifx{#1\undefined}
}%
\providecommand \@ifnum [1]{%
 \ifnum #1\expandafter \@firstoftwo
 \else \expandafter \@secondoftwo
 \fi
}%
\providecommand \@ifx [1]{%
 \ifx #1\expandafter \@firstoftwo
 \else \expandafter \@secondoftwo
 \fi
}%
\providecommand \natexlab [1]{#1}%
\providecommand \enquote  [1]{``#1''}%
\providecommand \bibnamefont  [1]{#1}%
\providecommand \bibfnamefont [1]{#1}%
\providecommand \citenamefont [1]{#1}%
\providecommand \href@noop [0]{\@secondoftwo}%
\providecommand \href [0]{\begingroup \@sanitize@url \@href}%
\providecommand \@href[1]{\@@startlink{#1}\@@href}%
\providecommand \@@href[1]{\endgroup#1\@@endlink}%
\providecommand \@sanitize@url [0]{\catcode `\\12\catcode `\$12\catcode
  `\&12\catcode `\#12\catcode `\^12\catcode `\_12\catcode `\%12\relax}%
\providecommand \@@startlink[1]{}%
\providecommand \@@endlink[0]{}%
\providecommand \url  [0]{\begingroup\@sanitize@url \@url }%
\providecommand \@url [1]{\endgroup\@href {#1}{\urlprefix }}%
\providecommand \urlprefix  [0]{URL }%
\providecommand \Eprint [0]{\href }%
\providecommand \doibase [0]{http://dx.doi.org/}%
\providecommand \selectlanguage [0]{\@gobble}%
\providecommand \bibinfo  [0]{\@secondoftwo}%
\providecommand \bibfield  [0]{\@secondoftwo}%
\providecommand \translation [1]{[#1]}%
\providecommand \BibitemOpen [0]{}%
\providecommand \bibitemStop [0]{}%
\providecommand \bibitemNoStop [0]{.\EOS\space}%
\providecommand \EOS [0]{\spacefactor3000\relax}%
\providecommand \BibitemShut  [1]{\csname bibitem#1\endcsname}%
\let\auto@bib@innerbib\@empty
%</preamble>
\bibitem [{\citenamefont {Aktay}\ \emph {et~al.}(2024)\citenamefont {Aktay},
  \citenamefont {Sander},\ and\ \citenamefont {Zochowski}}]{Aktay2024}%
  \BibitemOpen
  \bibfield  {author} {\bibinfo {author} {\bibfnamefont {Sinan}\ \bibnamefont
  {Aktay}}, \bibinfo {author} {\bibfnamefont {Leonard~M.}\ \bibnamefont
  {Sander}}, \ and\ \bibinfo {author} {\bibfnamefont {Michal}\ \bibnamefont
  {Zochowski}},\ }\bibfield  {title} {\enquote {\bibinfo {title}
  {{Neuromodulatory effects on synchrony and network reorganization in networks
  of coupled Kuramoto oscillators}},}\ }\href@noop {} {\bibfield  {journal}
  {\bibinfo  {journal} {Physical Review E}\ }\textbf {\bibinfo {volume}
  {110}},\ \bibinfo {pages} {044401} (\bibinfo {year} {2024})}\BibitemShut
  {NoStop}%
\bibitem [{\citenamefont {Sakaguchi}(1988)}]{Sakaguchi1988}%
  \BibitemOpen
  \bibfield  {author} {\bibinfo {author} {\bibfnamefont {Hidetsugu}\
  \bibnamefont {Sakaguchi}},\ }\bibfield  {title} {\enquote {\bibinfo {title}
  {{Cooperative Phenomena in Coupled Oscillator Systems under External
  Fields}},}\ }\href@noop {} {\bibfield  {journal} {\bibinfo  {journal}
  {Progress of Theoretical Physics}\ }\textbf {\bibinfo {volume} {79}},\
  \bibinfo {pages} {39--46} (\bibinfo {year} {1988})}\BibitemShut {NoStop}%
\bibitem [{\citenamefont {Kuramoto}(1984)}]{Kuramoto1984}%
  \BibitemOpen
  \bibfield  {author} {\bibinfo {author} {\bibfnamefont {Yoshiki}\ \bibnamefont
  {Kuramoto}},\ }\bibfield  {title} {\enquote {\bibinfo {title} {{Chemical
  Oscillations, Waves, and Turbulence}},}\ }\href@noop {} {\bibfield  {journal}
  {\bibinfo  {journal} {Springer Series in Synergetics}\ } (\bibinfo {year}
  {1984})}\BibitemShut {NoStop}%
\bibitem [{\citenamefont {Acebrón}\ \emph {et~al.}(2005)\citenamefont
  {Acebrón}, \citenamefont {Bonilla}, \citenamefont {Vicente}, \citenamefont
  {Ritort},\ and\ \citenamefont {Spigler}}]{Acebron2005}%
  \BibitemOpen
  \bibfield  {author} {\bibinfo {author} {\bibfnamefont {Juan~A.}\ \bibnamefont
  {Acebrón}}, \bibinfo {author} {\bibfnamefont {L.~L.}\ \bibnamefont
  {Bonilla}}, \bibinfo {author} {\bibfnamefont {Conrad J.~Pérez}\ \bibnamefont
  {Vicente}}, \bibinfo {author} {\bibfnamefont {Félix}\ \bibnamefont
  {Ritort}}, \ and\ \bibinfo {author} {\bibfnamefont {Renato}\ \bibnamefont
  {Spigler}},\ }\bibfield  {title} {\enquote {\bibinfo {title} {{The Kuramoto
  model: A simple paradigm for synchronization phenomena}},}\ }\href {\doibase
  10.1103/revmodphys.77.137} {\bibfield  {journal} {\bibinfo  {journal}
  {Reviews of Modern Physics}\ }\textbf {\bibinfo {volume} {77}},\ \bibinfo
  {pages} {137--185} (\bibinfo {year} {2005})}\BibitemShut {NoStop}%
\bibitem [{\citenamefont {Bi}\ and\ \citenamefont {Poo}(2001)}]{Bi2001}%
  \BibitemOpen
  \bibfield  {author} {\bibinfo {author} {\bibfnamefont {Guo-qiang}\
  \bibnamefont {Bi}}\ and\ \bibinfo {author} {\bibfnamefont {Mu-ming}\
  \bibnamefont {Poo}},\ }\bibfield  {title} {\enquote {\bibinfo {title}
  {{SYNAPTIC MODIFICATION BY CORRELATED ACTIVITY: Hebb's Postulate
  Revisited}},}\ }\href {\doibase 10.1146/annurev.neuro.24.1.139} {\bibfield
  {journal} {\bibinfo  {journal} {Annual Review of Neuroscience}\ }\textbf
  {\bibinfo {volume} {24}},\ \bibinfo {pages} {139--166} (\bibinfo {year}
  {2001})}\BibitemShut {NoStop}%
\bibitem [{\citenamefont {Maistrenko}\ \emph {et~al.}(2007)\citenamefont
  {Maistrenko}, \citenamefont {Lysyansky}, \citenamefont {Hauptmann},
  \citenamefont {Burylko},\ and\ \citenamefont {Tass}}]{Maistrenko2007}%
  \BibitemOpen
  \bibfield  {author} {\bibinfo {author} {\bibfnamefont {Yuri~L.}\ \bibnamefont
  {Maistrenko}}, \bibinfo {author} {\bibfnamefont {Borys}\ \bibnamefont
  {Lysyansky}}, \bibinfo {author} {\bibfnamefont {Christian}\ \bibnamefont
  {Hauptmann}}, \bibinfo {author} {\bibfnamefont {Oleksandr}\ \bibnamefont
  {Burylko}}, \ and\ \bibinfo {author} {\bibfnamefont {Peter~A.}\ \bibnamefont
  {Tass}},\ }\bibfield  {title} {\enquote {\bibinfo {title} {{Multistability in
  the Kuramoto model with synaptic plasticity}},}\ }\href@noop {} {\bibfield
  {journal} {\bibinfo  {journal} {Physical Review E}\ }\textbf {\bibinfo
  {volume} {75}},\ \bibinfo {pages} {066207} (\bibinfo {year}
  {2007})}\BibitemShut {NoStop}%
\bibitem [{\citenamefont {Berner}\ \emph {et~al.}(2021)\citenamefont {Berner},
  \citenamefont {Yanchuk}, \citenamefont {Maistrenko},\ and\ \citenamefont
  {Schöll}}]{Berner2021}%
  \BibitemOpen
  \bibfield  {author} {\bibinfo {author} {\bibfnamefont {Rico}\ \bibnamefont
  {Berner}}, \bibinfo {author} {\bibfnamefont {Serhiy}\ \bibnamefont
  {Yanchuk}}, \bibinfo {author} {\bibfnamefont {Yuri}\ \bibnamefont
  {Maistrenko}}, \ and\ \bibinfo {author} {\bibfnamefont {Eckehard}\
  \bibnamefont {Schöll}},\ }\bibfield  {title} {\enquote {\bibinfo {title}
  {{Generalized splay states in phase oscillator networks}},}\ }\href@noop {}
  {\bibfield  {journal} {\bibinfo  {journal} {Chaos: An Interdisciplinary
  Journal of Nonlinear Science}\ }\textbf {\bibinfo {volume} {31}} (\bibinfo
  {year} {2021})}\BibitemShut {NoStop}%
\end{thebibliography}%

\end{document}